\documentclass[aps,prl,twocolumn,superscriptaddress,letterpaper]{revtex4-2}

\usepackage{graphicx}
\usepackage{bm}
\usepackage{amssymb}
\usepackage{color}
\usepackage{amsmath}
\usepackage[colorlinks=true,linkcolor=blue,anchorcolor=black,citecolor=blue,filecolor=black,menucolor=black,runcolor=black,urlcolor=blue]{hyperref}
\usepackage{paracol}
\usepackage[many]{tcolorbox}

\definecolor{boxcolor}{RGB}{228,228,223}

\NewTColorBox{NewBox}{ s O{t} }{%
  floatplacement={#2},
  IfBooleanTF={#1}{float*,width=\textwidth}{float},
  colframe=white,colback=boxcolor,
  }

\begin{document}

\title{Topological acoustics}

\author{Haoran Xue}
\email{haoran001@e.ntu.edu.sg}
\affiliation{Division of Physics and Applied Physics, School of Physical and Mathematical Sciences, Nanyang Technological University,
Singapore 637371, Singapore}

\author{Yihao Yang}
\email{yangyihao@zju.edu.cn}
\affiliation{Interdisciplinary Center for Quantum Information, State Key Laboratory of Modern Optical Instrumentation, College of Information Science and Electronic Engineering, ZJU-Hangzhou Global Science and Technology Innovation Center, Key Lab. of Advanced Micro/Nano Electronic Devices \& Smart Systems of Zhejiang,  ZJU-UIUC Institute, Zhejiang University, Hangzhou 310027, China}

\author{Baile Zhang}
\email{blzhang@ntu.edu.sg}
\affiliation{Division of Physics and Applied Physics, School of Physical and Mathematical Sciences, Nanyang Technological University,
Singapore 637371, Singapore}
\affiliation{Centre for Disruptive Photonic Technologies, Nanyang Technological University,
Singapore 637371, Singapore}

\begin{abstract}
Topological acoustics is an emerging field that lies at the intersection of condensed matter physics, mechanical structural design and acoustics engineering. It explores the design and construction of novel artificial structures, such as acoustic metamaterials and phononic crystals, to manipulate sound waves robustly, taking advantage of topological protection. Early work on topological acoustics was limited to duplicating topological phases that have been understood in condensed matter systems, but recent advances have shifted the paradigm to exploring novel topological concepts that are difficult to realize in other physical systems, such as various topological semimetal phases, and topological phases associated with Floquet engineering, fragile topology, non-Hermiticity and synthetic dimensions. These developments demonstrate the  unique advantages of topological acoustic systems and their role in developing topological physics. In this Review, we survey the fundamental mechanisms, basic designs and practical realizations of topological phases in acoustic systems, and provide an overview of future directions and potential applications.     
\end{abstract}

\maketitle

Sound and vibrations are ubiquitous. However, it is still challenging to fully control them. A recent example is the 2011 Fukushima disaster, in which the seismic waves did not initially damage the highest-grade anti-seismic buildings, but destroyed the external power supply to the nuclear plant, triggering a series of reactions and eventually leading to the disaster \cite{kurokawa2013}. Another example is the low-frequency noise pollution that is a worldwide issue \cite{leventhall2004}, because it  can easily penetrate into most buildings. At the same time, the use of acoustic waves has become more and more important in modern technologies. For example, acoustic waves are used in biomedical microfluidic devices as a “tweezer” to trap and manipulate particles and cells \cite{ozcelik2018}. Moreover, acoustic waves can serve as information carriers that bridge the gap between electronics and photonics \cite{balram2016}; the frequencies of on-chip acoustic waves are close to those of central processing units and wireless communication systems, while their wavelengths are comparable to those of light. More recent studies have also demonstrated an acoustic approach to the manipulation of quantum information, for example, by coupling acoustic modes to a superconducting qubit \cite{chu2017}. The ability to manipulate acoustic waves in a controlled way with great robustness is hence strongly desirable. 

Robust properties, such as immunity to scattering caused by disorder or defects, can be found in phases of topological quantum materials (the representative topological phases are illustrated in {\bf Figs.~\ref{fig1}a-h)} \cite{hasan2010, qi2011, armitage2018}. Typical examples are topological insulators (TIs), including materials hosting the quantum Hall (QH) effect, which are insulating in the bulk, but conduct electricity at their boundaries owing to the existence of topological boundary states that are robust against backscattering \cite{hasan2010, qi2011}. Since 2015, the discovery of topological semimetals has extended topological phases to gapless systems \cite{armitage2018}. TIs and topological semimetals represent a revolution brought by the introduction of the concept of topology into condensed matter physics. 

A similar revolution can be envisioned in acoustics. In the past three decades, artificially structured materials have been engineered to control acoustic waves with tremendous developments in science and technology \cite{maldovan2013}. The concept of an acoustic or phononic bandgap was proposed in the 1990s \cite{kushwaha1993, sigalas1993}, laying the foundation for the field of phononic crystals for acoustic waves, similar to photonic crystals for light. Since early this century, deep subwavelength building blocks, or “meta-atoms”, have been engineered and incorporated in phononic crystals, giving rise to exciting discoveries, including negative refraction and invisibility cloaking for acoustic waves \cite{cummer2016}. Although there is still some debate in terminology, state-of-the-art artificially engineered acoustic structures are widely referred to as acoustic metamaterials and constitute a broad and independent field. Despite these remarkable developments, the concept of topology has  been introduced for acoustics and acoustic metamaterials only a few years ago.

This does not mean it is too late for topological acoustics to catch up with the frontiers of topological physics. On the contrary, we have witnessed the extremely rapid growth of this field ({\bf Fig.~\ref{fig1}l}). In the early years, topological acoustics borrowed ideas already understood in condensed matter systems \cite{ma2019, zhang2018}. This rapid growth has quickly caught up the pace of condensed matter developments, thanks to the relative ease of engineering scatterers and coupling in acoustic metamaterials. Nowadays, topological acoustics has become a field that realizes new topological phases beyond those that can be obtained in real materials, and offers novel ways to control the propagation of sound.

In this Review, we provide an overview of the developments in topological acoustics. We focus on acoustic waves in fluids, including air, where most studies have been performed. Elastic waves are covered in other recent reviews~\cite{huber2016, ma2019}. The first and second sections review the acoustic analogues of the QH topological phase and time-reversal ($\mathcal{T}$)-invariant topological insulating phases, including the quantum spin Hall (QSH) phase and valley Hall  phases ({\bf Figs.~\ref{fig1}a-d}). The third section surveys acoustic realizations of topological phases with quantized dipole and multipole moments, which are closely related to the recently discovered higher-order TIs (HOTIs; {\bf  Figs.~\ref{fig1}e-g}). The fourth section introduces  work on acoustic topological semimetals ({\bf Fig.~\ref{fig1}h}). In the fifth section, we highlight several emerging topological phases with novel degrees of freedom such as Floquet-type time modulation, fragile bands, non-Hermiticity and synthetic dimensions ({\bf Figs.~\ref{fig1}i-k}). Finally, we present an outlook on possible future directions for the field.
\bigskip

\begin{large}  
\noindent{\textbf{Effective magnetic fields and acoustic quantum Hall insulators}}
\end{large} 
\smallskip

The QH effect was the first phenomenon observed to manifest nontrivial band topology. It was first discovered in a 2D electron gas under a strong out-of-plane magnetic field, in which the Hall conductance takes integer values in units of $e^2/h$ \cite{klitzing1980}. This phenomenon was soon explained by David Thouless and co-workers, who revealed that the quantized Hall conductance is related to a topological invariant, later called the  Chern number, associated with the bulk bands \cite{thouless1982}. The Chern number is defined through the integration of the Berry curvature in the Brillouin zone (BZ) as
\begin{equation}
\mathcal{C}_n=\frac{1}{2\pi}\int_{\text{BZ}}\Omega_n(\bm{k})d^2\bm{k},
\end{equation}
where 
\begin{equation}
\Omega_n(\bm{k})=i(\langle\partial_{k_x}u_n(\bm{k})|\partial_{k_y}u_n(\bm{k})\rangle-\langle\partial_{k_y}u_n(\bm{k})|\partial_{k_x}u_n(\bm{k})\rangle)
\end{equation}
is the nonvanishing component of the Berry curvature, with $|u_n(\bm{k})\rangle$ the periodic part of the Bloch wavefunction of the $n$th band. The Chern number is necessarily integer-valued, counting the net number of chiral edge modes that enter and leave a given band. The chiral edge modes, which are responsible for the quantized Hall conductance, propagate along the edges unidirectionally, and are robust against gap-preserving disorder and defects ({\bf Fig.~\ref{fig1}a}). 

\begin{NewBox}*[t]
Box 1 $|$ \textbf{Analogous magnetic effects for sound}

\vskip\medskipamount 
\leaders\vrule width \textwidth\vskip0.4pt 
\vskip\medskipamount 

Sound waves are intrinsically inert to magnetic fields. However, one can still construct effective magnetic fields for sound and realize analogous magnetic effects in acoustic systems \cite{nassar2020}. As a first example, we discuss the realization of the Aharonov-Bohm (AB) effect \cite{aharonov1959} in acoustics. The AB effect occurs when electrons travel outside an infinitely long solenoid carrying a current (panel a). Although the magnetic field is entirely confined inside the solenoid, the nonzero vector potential outside modifies the phase of the electrons’ wavefunctions and leads to interference phenomena. This effect can be observed in water waves, as was demonstrated by Michael Berry and co-workers \cite{berry1980}. In their experiment, an irrotational vortex (panel b) played the role of the solenoid and the velocity field, which is curl-free, served as an effective vector potential. In such a setting, a wavefront dislocation is clearly observed when the surface water wave passes through the vortex from left to right (panel c). Following this idea,  an analog of the AB effect in acoustics was demonstrated by constructing a vorticity filament and letting an acoustic plane wave pass through it \cite{roux1997}. Wavefront dislocation was observed after the wave passed through the vorticity filament. 
\medskip

The second example is an acoustic analog of the Zeeman effect in a ring cavity with circulating airflow \cite{fleury2014}. When there is no airflow, the cavity supports two degenerate counterpropagating modes. When an airflow is applied, the two counterpropagating modes effectively circulate with different velocities and have different eigenfrequencies. Thus, airflow lifts the degeneracy of the two counterpropagating modes, in analog to the energy splitting of electronic orbitals induced by a magnetic field in the Zeeman effect.\\

Panel c adapted from \cite{berry1980} (Berry et al.).

\bigskip
\begin{center}
\includegraphics[width=0.9\textwidth]{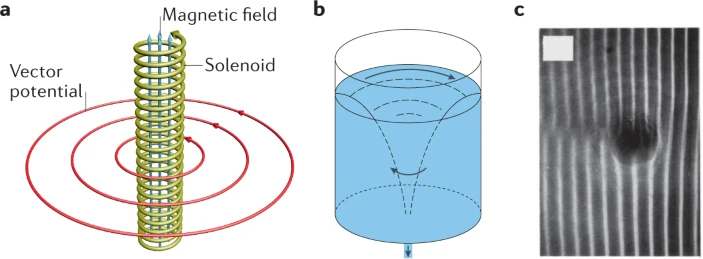}
\end{center}
\end{NewBox}

To realize a QH phase, it is necessary to break $\mathcal{T}$ symmetry. This is achievable in electronic systems by applying a magnetic field. In acoustic systems, one has to engineer an effective magnetic field instead. In fact, early studies showed that analogous magnetic effects can take place in acoustic systems with moving media ({\bf Box 1}) \cite{roux1997, fleury2014}. In 2015, several proposals for realizing acoustic QH insulators were put forward \cite{yang2015, ni2015, khanikaev2015}. The basic idea was to arrange circulating flows into periodic settings, forming a $\mathcal{T}$-broken acoustic lattice. For example, a triangular lattice consisting of rotating rods was designed as shown in {\bf Fig.~\ref{fig2}a}. The blue regions are filled with fluids that perform a circulatory motion due to the rods’ rotation. The remaining area is filled with a stationary fluid (such as air). The circulating flows induce a nontrivial bandgap transversed by chiral edge modes ({\bf  Figs.~\ref{fig2}b and \ref{fig2}c}). These chiral edges modes, as expected, travel unidirectionally along the lattice edges and are robust against disorder and defects ({\bf  Fig.~\ref{fig2}d}).

To gain more insights into why QH insulators can be realized in such a lattice, let us consider the following master equation for sound with negligible viscosity and heat flow \cite{brekhovskikh1999}:
\begin{equation}
\frac{1}{\rho}\nabla\cdot\rho\nabla\phi-(\partial_t+\bm{v}\cdot\nabla)\frac{1}{c^2}(\partial_t+\bm{v}\cdot\nabla)\phi=0.
\end{equation}
Here, $\rho$ is the fluid density, $c$ is the sound speed, and $\bm{v}$ is the fluid velocity. By looking for time-harmonic solutions with angular frequency $w$ and letting $\Psi=\sqrt{\rho}\phi$, the master equation can be written in a compact way as \cite{yang2015}
\begin{equation}
[(\nabla-i\bm{A}_{\text{eff}})+V]\Psi=0,
\end{equation}
where $\bm{A}_{\text{eff}}=-\omega\bm{v}/c^2$ is the effective vector potential induced by a nonzero $\bm{v}$. Thus, the circulating flows induce an effective magnetic field for sound waves. For the lattice given in {\bf Fig.~\ref{fig2}a}, both positive and negative effective magnetic fluxes are induced, leading to zero net magnetic flux within one unit cell \cite{yang2015}, similar to anomalous QH insulators \cite{haldane1988}. 

Acoustic QH phases can also be implemented in other lattice geometries with flows, such as honeycomb \cite{ni2015, khanikaev2015} and square lattices \cite{chen2016}. Moreover, it is possible to construct an acoustic QH phase in active-liquid metamaterials, where the flow can be generated by self-propelled particles without using external drives \cite{souslov2017}. QH physics can also be studied in lattices with synthetic dimensions, as discussed later. On the experimental side, realizing an acoustic QH insulator is quite challenging owing to undesired effects such as non-synchronous rotation and flow instabilities. To overcome these issues, an acoustic ring resonator lattice was designed with optimized structural parameters and a high-order mode with high quality factor, which largely reduces the required airflow speed \cite{ding2019}. Based on this design ({\bf Fig.~\ref{fig2}e}), acoustic chiral edge modes were successfully observed \cite{ding2019}.

Finally, it is worth mentioning that in $\mathcal{T}$-invariant acoustic systems, although the genuine QH phase cannot exist, one can still construct so-called pseudo magnetic fields and realize some analogous QH effects via proper structural engineering \cite{yang2017, wen2019}. This approach is in the same spirit as strain engineering in graphene \cite{guinea2010}, where mechanical deformations couple to Dirac cones as gauge fields and lead to the formation of pseudo Landau levels connected by in-gap edge states (note that these edge states are not unidirectional). Furthermore, such pseudo magnetic fields can also be induced in acoustic Weyl crystals, leading to chiral pseudo Landau levels \cite{peri2019} (as discussed in the section on acoustic topological semimetals).

\bigskip
\begin{large}
\noindent{\textbf{Pseudospins and $\mathcal{T}$-preserved topological phases in acoustics}}
\end{large}
\smallskip

In acoustic QH insulators, fluid flow is usually required to break $\mathcal{T}$ symmetry, which makes experimental realizations challenging. In this section, we discuss approaches to realize $\mathcal{T}$-invariant acoustic topological insulating phases, or the so-called acoustic QSH and acoustic valley Hall insulators, by exploiting spin and valley degrees of freedom.

\bigskip
\noindent\textbf{Acoustic analogues of quantum spin Hall insulators}
\smallskip

The QSH insulator \cite{kane2005a, kane2005b, bernevig2006}, which can be regarded as a system simultaneously supporting two time-reversed copies of the QH insulator, hosts so-called helical edge states. The term ``helical" refers to the fact that opposite spins counterpropagate along the edge \cite{wu2006}, in contrast to chiral edge states that propagate unidirectionally ({\bf Figs.~\ref{fig1}a, b}). In a QSH insulator, the Chern number for each spin is nonzero, despite the vanishing total Chern number ($C_{\text{up}}=-C_{\text{down}}\neq0$). Unlike the QH phase, a QSH insulator does not break $\mathcal{T}$ symmetry but is instead protected by $\mathcal{T}$ symmetry. This is because $\mathcal{T}$ symmetry for an electron satisfies $\mathcal{T}^2=-1$, which enables Kramers doublets, where opposite spins are degenerate at $\mathcal{T}$-invariant momenta. It is thus possible to achieve gapless edge states robust against $\mathcal{T}$-preserving perturbations. However, for sound, which carries intrinsic spin-0, $\mathcal{T}$ symmetry satisfies $\mathcal{T}^2=1$. Such a fundamental difference poses a challenge to the implementation of an acoustic analogue.

A common strategy to overcome this issue is to modify $\mathcal{T}$ into $U\mathcal{T}$ ($U$ is an operator usually generated from a lattice symmetry), such that $(U\mathcal{T})^2=-1$. Subsequently, fermion-like pseudospins and the corresponding Kramers-like degeneracy can be constructed. One way to do so is based on tuning the accidental double Dirac cone in graphene-like lattices ({\bf Fig.~\ref{fig2}f}) \cite{he2016, mei2016}. By tuning the structural parameters, a band inversion associated with a topological phase transition can occur at the $\Gamma$ point, as illustrated in {\bf Fig.~\ref{fig2}g}. At the interface between a topological acoustic crystal and a trivial one, one pair of gapless modes emerge ({\bf Fig.~\ref{fig2}h}). These interface modes are indeed counterpropagating helical modes ({\bf Fig.~\ref{fig2}i}) \cite{he2016}. A similar approach to achieving an acoustic QSH insulator is via a BZ folding mechanism \cite{zhang2017, yves2017, xia2017, deng2017}. The starting point is an acoustic crystal featuring a pair of Dirac points at the K and $\text{K}’ $ corners. By properly choosing a supercell, the Dirac points are folded to the $\Gamma$ point. Then a  topological phase transition takes place via shrinking or expanding the supercell. In these systems, pseudospins are formed by hybridization of the orbital modes (note the pseudospins are only well-defined around the $\Gamma$ point), and the condition $(U\mathcal{T})^2=-1$ can be fulfilled by using a $U$ generated from the $C_{6}$ symmetry \cite{wu2015}. However, at the interface, the protective symmetry can be slightly broken, resulting in a tiny gap in the edge dispersion (not visible in {\bf Fig.~\ref{fig2}h}).

Acoustic QSH insulators can also be implemented using other pseudospin degrees of freedom. For example, a double-layer acoustic crystal was realized with a layer pseudospin degree of freedom, where the interlayer coupling is analogous to the spin-orbit coupling \cite{deng2020}. Remarkably, this design shows gapless helical edge modes at the external boundaries rather than at the domain walls or interfaces as in the previous models \cite{he2016, mei2016, zhang2017, yves2017, xia2017, deng2017}. As the bandgap of an acoustic QSH insulator is usually narrow, a recent work adopted topology optimization to maximize the bandgap width \cite{christiansen2019}. A reconfigurable acoustic QSH insulator was also experimentally demonstrated \cite{xia2018}. Furthermore, the QSH phase can be extended to 3D systems\cite{fu2007}. Recently, an acoustic analog of a 3D QSH insulator with gapless surface Dirac cones was realized \cite{he2020}.

\bigskip
\noindent\textbf{Acoustic valley Hall insulators}
\smallskip

Initially inspired by advances in 2D materials research \cite{schaibley2016}, the valley degree of freedom has been introduced to acoustics \cite{lu2016, lu2017}. A valley is a local extremum in the band structure, which is generally located at high-symmetry momenta. Acoustic valley Hall insulators are usually obtained by gapping out Dirac points through the breaking of inversion ($\mathcal{P}$) or mirror symmetry. Around the valleys, so-called valley-contrasting physics can occurr, that is, opposite orbital magnetic moment and Berry curvatures can be realized at different valleys \cite{xiao2007}. This leads to various valley-dependent phenomena; in particular, angular-momentum-carrying valley pseudospins and valley kink states are widely studied in acoustics.

A typical acoustic valley Hall insulator is shown in {\bf Fig.~\ref{fig2}j}, and features triangular scatterers placed in a triangular lattice with a background of air. When the orientation angle $\theta$ of scatterers is 0$^\circ$, the acoustic crystal has mirror symmetry, and a pair of Dirac points appear at the BZ corners (K and $\text{K}’ $ points). Slightly altering $\theta$ breaks mirror symmetry, lowering the symmetry from ${C}_{3v}$ to ${C}_{3}$ at the K and $\text{K}’ $ points and lifting the degeneracy of the Dirac points \cite{lu2014b, lu2016, lu2017}. The Bloch modes at the K and $\text{K}’ $ valleys exhibit vortices with either left-handed circular polarization or right-handed circular polarization \cite{lu2016} ({\bf Fig.~\ref{fig2}k}). These valley vortex states offer controlled propagation of bulk modes, such as valley-chirality locked beam splitting \cite{lu2016, ye2017}. Furthermore, at a domain wall, also called a kink, between two acoustic valley Hall lattices with opposite $\theta$, so-called valley kink states emerge, with propagation directions locked to the valley degree of freedom ({\bf Fig.~\ref{fig2}l}). These kink states are characterized by a half-integer topological invariant called the valley Chern number, which is the integration of Berry curvature around one valley. It was experimentally demonstrated that the kink states can pass through 120$^\circ$ sharp corners with negligible scattering ({\bf Fig.~\ref{fig2}m}). In addition, studies have also demonstrated the robustness of valley kink states against other types of path bends \cite{lu2017, wu2017, zhang2019e} and certain types of disorder \cite{orazbayev2019}. These properties can be used to devise devices like robust acoustic delay lines \cite{zhang2018c}. Another interesting property of the valley kink states is that at zigzag terminations, the valley kink states can outcouple to the surrounding background with 100\% efficiency \cite{ma2016, gao2018, jia2021}, which is useful for designing directional acoustic antennas \cite{zhang2018b}.

Valley Hall phases can be further enriched in multilayer structures. Researchers have found that a bilayer acoustic valley Hall insulator can exhibit layer-mixed and layer-polarized valley kink states \cite{lu2018, zhu2021}. By adding a finite-width acoustic crystal featuring Dirac points between two topologically distinct acoustic valley Hall insulators, a topological valley-locked large-area waveguide was demonstrated \cite{wang2020}. Such a waveguide can be used for robust high-energy-capacity transport and valley-locked converging of sound. Moreover, by introducing on-site gain and loss to an acoustic valley Hall insulator, the valley vortex states and kink states can be selectively attenuated or amplified \cite{wang2018}.

\bigskip
\begin{large}
\noindent\textbf{Acoustic topological phases with quantized dipole and multipole moments}
\end{large}
\smallskip

In the previous two sections, we have discussed the QH, QSH and valley Hall phases in acoustics. These topological phases are characterized by the Chern number and its derivatives, like the spin and valley Chern numbers, which are integrations of the Berry curvature. In this section, we turn our focus to another large class of topological phases that are characterized by a quantized dipole moment, that is, the integration of the Berry connection, and by its generalizations, such as quantized quadrupole and octupole moments.

\begin{NewBox}
Box 2 $|$ \textbf{Acoustic implementation of tight-binding lattices}

\vskip\medskipamount 
\leaders\vrule width \textwidth\vskip0.4pt 
\vskip\medskipamount 
\nointerlineskip

In acoustic systems, many topological lattice models can be implemented in a straightforward manner. The building blocks for implementing tight-binding lattices in acoustics are acoustic cavities and tubes, playing the roles of atoms and hoppings, respectively (panel \textbf{a}). The cavities support different resonance modes at different frequencies. Without the tubes, the system is described by a diagonal matrix $H_0$ whose elements are the eigenfrequencies. When the tubes are introduced, the lattice is described by the matrix $H_0+H_1$, where $H_1$ accounts for the effects introduced by the tubes \cite{matlack2018}. These are couplings between the modes of interest, couplings between the modes of interest and other modes (this indicates $H_1$ is in general not block-diagonal), and on-site frequencies shifts of the resonance modes (this indicates $H_1$ may contain diagonal elements). The latter two effects are undesired, and can be made very weak by using weak couplings and choosing a resonance mode that is well-separated from other modes.
\medskip

Many degrees of freedom can be tuned in such coupled acoustic resonator lattices. For example, the on-site frequency ($f_0$) can be controlled by tuning the geometry of the cavity, and the coupling strength ($\gamma_{1,2}$) and sign can be controlled by the size and location of the tubes (panel \textbf{b}) \cite{ni2020, xue2020, qi2020}. It is also possible to realize non-Hermitian terms such as on-site gain and loss or nonreciprocal hoppings by introducing extra absorbing materials or active components \cite{ding2016, gao2020, gao2021, zhang2021b}.

\includegraphics[width=\columnwidth]{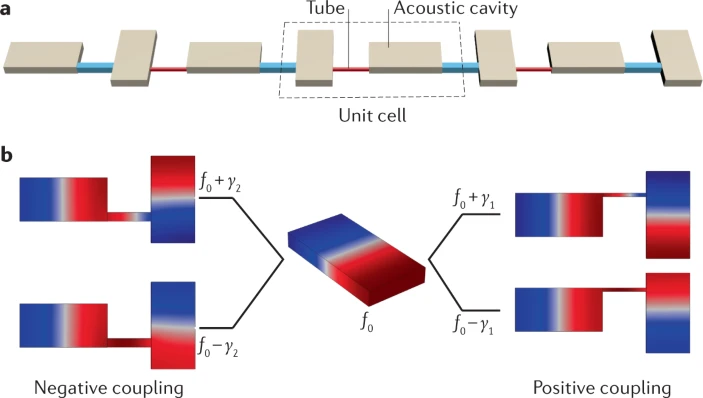}
\end{NewBox}

\bigskip
\noindent{\textbf{1D acoustic topological phases}}
\smallskip

In this subsection, we briefly discuss 1D acoustic topological phases~\cite{ma2019}. Acoustic topological phases in 1D are characterized by quantized bulk dipole polarization. In the modern theory of polarization \cite{king1993, vanderbilt1993}, the bulk dipole moment is formulated through the Berry phase \cite{berry1984} as
\begin{equation}
P=\frac{1}{2\pi}\int_k^{k+2\pi}\text{Tr}[A_k]d\bm{k},
\end{equation}
where $[A_k]^{mn}=i\langle u_k^m|\partial_k|u_k^n\rangle$ is the Berry connection. Here, $|u_k^{m/n}\rangle$ is the periodic part of the Bloch wavefunction and $m$ and $n$ label the occupied bands (or equivalently, bands below the bandgap of interest). The dipole moment corresponds to the Wannier center (WC) and in general can take any value within one unit cell. Importantly, the dipole moment can be quantized by symmetries such as mirror and chiral symmetries, making it eligible as a topological invariant. Note that the dipole moment by definition is a gauge-variant quantity that depends on the choice of a unit cell. Nevertheless, the WC positions are unambiguous. When a 1D system has a nontrivial dipole moment, there are fractional charges at boundaries. The quantized dipole moment, as well as the fractional boundary charge, cannot be removed by perturbations that preserve both the protective symmetry and the bandgap. In realistic acoustic systems, although there is no actual polarization, because there are no charged particles like electrons, the above picture still applies. The dipole polarization can be understood as the WC, and the fractional boundary charge can be interpreted as a “fractional boundary anomaly” \cite{peterson2020}. 

The first demonstration of a 1D acoustic topological phase utilized an inversion-symmetric lattice ({\bf Fig.~\ref{fig3}a}) \cite{xiao2015a}. In this setup, two lattices with distinct dipole moment (or equivalently, Zak phase \cite{zak1989}) are adjacent, and generate an interfacial topological state. A 1D acoustic topological system can also be straightforwardly constructed by implementing certain tight-binding models using coupled acoustic resonators ({\bf Box 2}) \cite{yang2016, xiao2017, li2018}, such as the Su-Schrieffer-Heeger (SSH) model \cite{su1979}. These 1D acoustic topological systems provide platforms that give direct access to bulk quantities such as the Zak phase \cite{xiao2015a}. The topological boundary modes can be used to construct topological Fano resonances \cite{zangeneh2019} and achieve robust analog signal processing \cite{zangeneh2019b}.

\bigskip
\noindent\textbf{Acoustic higher-order topological insulating phases}
\smallskip

HOTIs are recently discovered topological materials that host topological modes at corners or hinges ({\bf {}Figs.~\ref{fig1}e-g}) \cite{xie2021}. Acoustic systems have served as important and versatile experimental platforms for demonstrating the physics of HOTIs. In particular, various types of HOTIs with corner modes have been realized in acoustics \cite{xue2019a, ni2019, zhang2019a, xue2019b, zhang2019b, zhang2019c, weiner2020, zhang2020, ni2020, xue2020, qi2020, yang2020, meng2020, zheng2020, yan2020, huang2022, zhang2021, yang2021, zheng2021}.

HOTIs with corner modes can be regarded, in general, as generalizations of 1D quantized dipole moments to higher dimensions. There are two types of such generalizations: first, the generalization of the quantized dipole moment in 1D systems to higher-dimensional systems \cite{ezawa2018a, ezawa2018b, xie2018, benalcazar2019}, and second, the generalization of quantized dipole moments to quantized multipole moments \cite{benalcazar2017a, benalcazar2017b}. In both cases, the band topology is captured by Berry-phase-type invariants, unlike in 2D QH and QSH TIs, whose topology is determined by the windings of the Berry phase. Such a difference reveals a fundamentally new notion of topology in HOTIs: the nontrivial topology is defined through the obstruction to moving the WC to the atomic site in the unit cell instead of the obstruction to finding a set of symmetric Wannier functions for the filled bands \cite{wieder2021}.

The first generalization is straightforward: the dipole moment in 2D and 3D is a vector that consists of the dipole polarization along each direction ($\bm{P}=(P_x, P_y, P_z)$). Similar to 1D cases, the dipole moment in 2D and 3D can be quantized by symmetries and thus can give rise to fractional charges at corners. Because the value of $\bm{P}$ gives the position of the WC, HOTIs with quantized dipole moment are also called Wannier-type HOTIs \cite{ezawa2018b}. A typical example of a Wannier-type HOTI is the kagome lattice with dimerized couplings \cite{ezawa2018a}, which was experimentally implemented in acoustics with coupled-resonator lattices \cite{xue2019a, ni2019} ({\bf Fig.~\ref{fig3}b}). Designs beyond tight-binding models have also been demonstrated \cite{zhang2019a}. In 3D, Wannier-type HOTIs have been implemented in a few acoustic systems, including the acoustic diamond lattice \cite{xue2019b}, the acoustic pyrochlore lattice ({\bf Fig.~\ref{fig3}c}) \cite{weiner2020} and an SSH-inspired acoustic cubic lattice \cite{zhang2019b}.

The second generalization leads to another type of HOTIs, namely multipole TIs. Multipole TIs have vanishing dipole moment but quantized multipole (such as quadrupole and octupole) moments, whose theoretical characterization involves novel techniques such as nested Wilson loops \cite{benalcazar2017a, benalcazar2017b} or many-body multipole operators \cite{kang2019, wheeler2019}. A pioneering work by Benalcazar, Bernevig and Hughes (BBH) first proposed a 2D lattice model for a quantized quadrupole TI and a 3D lattice model for a quantized octupole TI \cite{benalcazar2017a}. These models require negative couplings that are hard to realize in real materials but are feasible in acoustics. So far, both 2D and 3D BBH models have been successfully implemented in acoustics \cite{ni2020, xue2020, qi2020} ({\bf Fig.~\ref{fig3}d} left and {\bf Fig.~\ref{fig3}e}). These realizations utilize coupled acoustic cavities that host dipolar resonances and construct negative couplings by engineering the coupling channels ({\bf Box 2}). In acoustics, it is also possible to go beyond BBH models and realize multipole TIs without negative couplings. This was demonstrated by designing a non-symmorphic acoustic crystal ({\bf Fig.~\ref{fig3}d}, right) that realizes a quadrupole TI protected by two noncommutative glide symmetries \cite{zhang2020}. Furthermore, a 4D hexadecapolar TI was also experimentally studied in a 1D acoustic system via dimensional reduction \cite{chen2021b}.

It is worthwhile to point out that, although most studies have focused on corner modes, the decisive feature of higher-order band topology is the fractional corner charge. A recent study has shown that fractional corner charge can indeed be detected through accurate measurement of the local density of states in an electromagnetic system \cite{peterson2020}. It is still an open question whether similar measurements can be performed in acoustics. Apart from corner modes, 1D hinge modes have also been realized in 3D acoustic HOTIs \cite{he2020,wei2021b, du2022}.

\bigskip
\begin{large}
\noindent\textbf{Acoustic topological semimetals}
\end{large}
\smallskip

The topological properties of topological semimetals in condensed matter systems \cite{armitage2018, hasan2021} are defined by band degeneracies  where two or more bands intersect with each other in momentum space. Topological semimetals have also found their analogues in acoustics, which we refer to as acoustic topological semimetals. 

\bigskip
\noindent{\textbf{Acoustic Weyl crystals}}
\smallskip

Acoustic Weyl crystals \cite{xiao2015b, yang2016, li2018b, ge2018, xie2019b, huang2020b, zangeneh2020} host so-called Weyl points, where two bands linearly intersect in 3D momentum space ({\bf Fig.~\ref{fig1}h}), similar to Weyl semimetals in electronic materials. Weyl points are drains or sources of Berry flux and carry topological charges defined by Chern numbers. Excitations near the Weyl points are described by the Weyl Hamiltonian $H(\bm{k})=\bm{k}\cdot\bm{\sigma}$, with  $\bm{\sigma}$ the Pauli matrices. Because all three Pauli matrices are involved, Weyl points are very robust and can only be annihilated in pairs with opposite charges. Besides, according to the Nielsen-Ninomiya theorem \cite{nielsen1981}, the sum of topological charges of Weyl points in a crystal must be zero. An intriguing property of Weyl semimetals is that the isofrequency contours of surface states are {open} arcs, known as Fermi arcs \cite{wan2011}. These Fermi arcs connect the projections of two oppositely charged Weyl points ({\bf  Fig.~\ref{fig4}a}).

As the Berry curvature in momentum space is zero everywhere for a system with both  $\mathcal{P}$ and $\mathcal{T}$ symmetries \cite{xiao2010}, to achieve Weyl points in acoustics, either $\mathcal{P}$ or $\mathcal{T}$ should be broken. Because breaking $\mathcal{T}$ is very challenging in acoustics, all acoustic Weyl crystals realized so far have broken $\mathcal{P}$ and preserved $\mathcal{T}$. {\bf Fig.~\ref{fig4}b} shows the unit cell design of an acoustic Weyl crystal. For fixed $k_z$, the system can be viewed as the well-known Haldane lattice \cite{haldane1988}, with nonzero effective gauge flux generated by the chiral interlayer couplings \cite{xiao2015b}. Acoustic Weyl crystals are usually fabricated with the aid of 3D printing technology ({\bf Fig.~\ref{fig4}c}). Using direct acoustic measurements, the bulk dispersions of the Weyl points, as well as the acoustic Fermi arcs, can be observed \cite{li2018b} ({\bf  Fig.~\ref{fig4}d}).

Unlike condensed matter designs that must consider limitations of realistic materials such as their stability, acoustic systems can be built up almost at will. Besides, the probing frequency in acoustic systems is almost arbitrary, with no restriction as in condensed matter systems where probing must be near the Fermi level. Therefore, acoustic systems enable the observation of exotic phenomena previously inaccessible or extremely challenging in condensed matter systems. One example is the topological negative refraction of the Fermi arc surface states. By judiciously engineering the Fermi arcs on two adjacent surfaces, Fermi-arc surface states can be made to refract from one surface to the other without reflection, due to the non-closed nature of Fermi arcs ({\bf Fig.~\ref{fig4}e})\cite{he2018}. This is in sharp contrast to the conventional refraction: in the topological negative refraction, the incident and refracted beams are at the opposite sides of the normal and reflection is absent, whereas in nontopological normal refraction the two beams are on the same side of the normal, and reflection is usually unavoidable.

By designing an inhomogeneous acoustic Weyl crystal with the Weyl point positions continuously moving in one specific direction in 3D momentum space, a pseudo axial field can be created in acoustics, leading to the observation of chiral Landau levels \cite{peri2019} ({\bf Fig.~\ref{fig4}f}). Higher-order acoustic Weyl crystals exhibiting both Fermi arc states on 2D surfaces and hinge arc states at 1D hinges ({\bf  Fig.~\ref{fig4}g}) were also experimentally realized \cite{luo2021, wei2021}. Weyl points in these higher-order acoustic Weyl crystals can be viewed as transition points between Chern insulators and HOTIs parameterized by $k_z$.

\bigskip
\noindent{\textbf{Acoustic topological semimetals beyond conventional Weyl phases}}
\smallskip

As it turns out, there are various topological band degeneracies beyond the conventional Weyl point, which can be classified by the order of dispersion, the number of degenerate bands, and the dimensionality of the degeneracy ({\bf Fig.~\ref{fig1}h}). 

The 3D Dirac point is a four-fold linear point degeneracy, which can be viewed as an overlap of two Weyl points with opposite topological charges and can be described by an effective Hamiltonian $H(\bm{k})=[\bm{k}\cdot\bm{\sigma},0;0,-\bm{k}\cdot\bm{\sigma}]$ \cite{cheng2020, cai2020, xie2020, su2022}. Therefore, the 3D Dirac point carries a $Z_2$ topological charge. There are basically two categories of 3D Dirac points, one relying on band inversion \cite{xie2020} and the other stabilized by crystalline symmetries \cite{cheng2020, cai2020}. A special case of the latter was experimentally observed in acoustic crystals with space groups No.~206 and 230, which exhibit gapless quad-helicoid surface states stabilized by nonsymmorphic and $\mathcal{T}$ symmetries \cite{cheng2020, cai2020} ({\bf Fig.~\ref{fig5}a}). Note that though the surface states of the Dirac points have topological origin, they are not topologically protected in general and can be gapped by hybridizing two surface states with opposite pseudospins \cite{kargarian2016}. Furthermore, hinge states induced by higher-order band topology was recently discovered in acoustic Dirac crystals \cite{qiu2021, xia2022}.

The charge-2 Dirac point is also a four-fold linear point degeneracy, which, however, is a direct sum of two identical Weyl points and is described by the effective Hamiltonian $H(\bm{k})=[\bm{k}\cdot\bm{\sigma},0;0,\bm{k}\cdot\bm{\sigma}]$ \cite{yang2019}. It thus carries topological charge 2 and has double Fermi arcs ({\bf Fig.~\ref{fig5}b}). 

The band degeneracy can also be three-fold, exemplified by the spin-1 Weyl point \cite{yang2019}, which has two linearly dispersive bands and a flat band, labelled with Chern number $\pm2$ and 0, respectively. Unlike the spin-1/2 Weyl and Dirac quasiparticles, the excitations near the spin-1 Weyl point behave like spin-1 particles. The corresponding Hamiltonian is $H(\bm{k})=\bm{k}\cdot \bm{L}$, with $\bm{L}$ the spin-1 matrix representations \cite{yang2019}. Both the spin-1 Weyl point and the charge-2 Dirac point can be found in a chiral acoustic crystal without inversion or mirror symmetry (more specifically, an acoustic crystal with space group No.~198) and are stabilized by rotational, screw and $\mathcal{T}$ symmetries. Owing to the topological charge 2, the projected spin-1 Weyl point emanates two Fermi arcs ({\bf Fig.~\ref{fig5}b}), which connect the projection of the spin-1 Weyl point at the BZ centre and the charge-2 Dirac point at the BZ corner, forming a noncontractible loop that winds around the surface BZ torus \cite{yang2019}. These topological surface states exhibit topological negative refraction over all surfaces of the sample. Note that a spin-1 Weyl point is not necessarily paired with a charge-2 Dirac point. It can also be paired with another spin-1 Weyl point with opposite topological charge \cite{deng2020b}.

Topological degeneracies can also have quadratic or higher-order dispersions \cite{fang2012}. One example is the quadratic Weyl point, which usually has quadratic dispersion in two directions and linear dispersion in the third one. The quadratic Weyl point at $\mathcal{T}$-invariant momentum was experimentally realized in a chiral acoustic crystal with space group No.~181 and is enforced by the screw rotational and $\mathcal{T}$ symmetries \cite{he2020b}. Such a point degeneracy  carries topological charge 2 and exhibits two Fermi arcs \cite{he2020b} ({\bf Fig.~\ref{fig5}c}). 

The band degeneracies can also form 1D nodal lines \cite{deng2019, geng2019, qiu2019, xiao2019}. These nodal lines have various forms, such as chains, links, knots and rings. Besides, the nodal lines carry nontrivial Berry phase $\pi$, the same as 2D Dirac points. Acoustic nodal rings can be pinned at certain planes in 3D momentum space \cite{deng2019} ({\bf Fig.~\ref{fig5}d}) and exhibit topological flat drumhead states on the surfaces. Acoustic nodal lines winding around the BZ and the corresponding waterslide surface states that connect nodal lines with opposite chirality were also realized \cite{qiu2019}. Two intersecting nodal rings can form a nodal chain; interestingly, the Berry phase around the intersection point is zero. Nodal chains protected by nonsymmorphic and $\mathcal{T}$ symmetries were recently realized \cite{lu2020}. Nodal rings may also carry chiral topological charges in an acoustic crystal with broken inversion symmetry \cite{xiao2019}. 

Finally, the dimensionality of the band degeneracies can even be 2D. A 2D nodal surface stabilized by non-symmorphic lattice symmetry and $\mathcal{T}$ symmetry was realized by two groups independently \cite{yang2019b, xiao2020}. Interestingly, due to the broken inversion symmetry of the underlying acoustic crystal, the nodal surface carries a chiral charge +2. The nodal surface thus emanates two Fermi arcs connecting to two charge-1 Weyl points \cite{yang2019b} ({\bf  Fig.~\ref{fig5}e}).

\bigskip
\begin{large}
\noindent\textbf{Emerging topological phases in acoustics}
\end{large}
\smallskip

In this section, we review several emerging directions in topological acoustics, including Floquet, fragile and non-Hermitian topological phases, as well as topological phases with synthetic dimensions. 

\bigskip
\noindent{\textbf{Floquet topology}}
\smallskip

Apart from static systems, nontrivial band topology can also be found in systems with periodic modulation, also known as Floquet systems \cite{rudner2020}. A Floquet system is described by a time-dependent Hamiltonian $\hat{H}(t)=\hat{H}(T+t)$, with $T$ the driving period ({\bf Fig.~\ref{fig1}i}). The evolution of the system over a period is determined by the one-period evolution operator $\hat{U}(T)=\hat{T}\text{exp}[-i\int_{t_0}^{t_0+T}\hat{H}(t)dt]$ with $\hat{T}$ the time-ordering operator. The band structure, also known as the Floquet spectrum, can be obtained from $\hat{U}(T)\psi(t_0)=e^{-i\varepsilon T}\psi(t_0)$, where the quasi-energy $\varepsilon$ is an angular variable with period $2\pi/T$.

There are basically three approaches to realize Floquet acoustic TIs. The first is to apply temporal modulation to an acoustic crystal \cite{fleury2016, liu2019}. A time-dependent acoustic crystal with a hexagonal lattice of trimers was theoretically proposed \cite{fleury2016}. Each trimer consists of three coupled acoustic cavities, which possess acoustic capacitances that are periodically modulated in time in a rotating fashion ({\bf Fig.~\ref{fig6}a}). By doing so, the Floquet band structure opens topological bandgaps with nonzero Chern numbers, due to the broken $\mathcal{T}$ symmetry. This acoustic Floquet Chern insulator, similar to the Chern insulators in static systems, exhibits robust unidirectional edge states that can bypass defects, sharp corners and other types of disorder, and can be used for broadband acoustic diodes and topologically protected leaky antennas \cite{fleury2016}. 

The second approach is to map the periodically driven lattice to a scattering network, where a unitary scattering matrix plays the role of the Floquet evolution operator \cite{liang2013, pasek2014, gao2016}. As an external drive is not required, this mapping greatly facilitates experiments. In acoustics, this idea can be realized using 2D coupled metamaterial ring lattices \cite{peng2016, peng2017, wei2017}, where the clockwise/anticlockwise circulating mode in each ring can be viewed as pseudospin-up/pseudospin-down. By properly tuning the coupling strength between neighbouring rings, robust pseudospin-dependent topological edge states can be obtained ({\bf Fig.~\ref{fig6}b}). Such an acoustic lattice is also a realization of the so-called anomalous Floquet TI, whose topolgical edge states are even more robust than their counterparts in Chern insulators \cite{zhang2021d}, despite the vanishing Chern numbers in the bulk bands \cite{rudner2013}.

The third approach is to use a spatial dimension as an effective time dimension \cite{peng2018, peng2019, zhu2020}. With this approach, the dynamics of a $d$-dimensional Floquet system can be studied in a $(d+1)$-dimensional static system, provided there is negligible backscattering along the effective time axis. Based on this scheme, a 3D structure with couplings periodically modulated in the $z$ direction was experimentally demonstrated, showing broadband and low-loss effective chiral edge modes \cite{peng2019}. A 2D acoustic Floquet HOTI \cite{zhu2020} was also demonstrated following a similar strategy; its Floquet corner states have time-periodic evolution with a period longer than the underlying drive. This higher-order topological phase is anomalous in the sense that both dipole and quadrupole moments are vanishing.

\bigskip
\noindent{\textbf{Fragile topology}}
\smallskip

Fragile topology is a recently developed concept that describes a special type of topological matter \cite{po2018}. Unlike trivial bands, fragile topological bands do not admit a symmetric Wannier representation (one cannot adiabatically deform the system to a trivial atomic insulator while preserving the symmetries without closing the bandgap). However,  fragile topological bands are also not like conventional stable topological bands in the sense that the obstruction to the atomic limit can be removed by adding fully trivial bands. Thus, the boundary modes in a fragile TI are removable by symmetry-preserving perturbations. 

It is important to point out that fragile topology is commonly present in realistic systems. For example, fragile topology exists in twisted bilayer graphene that hosts superconducting states at the “magic angle” \cite{song2019}. In classical systems, including acoustics, many classical analogues of the QSH insulators, which have been found to support robust boundary propagation in experiments, were also proven to possess fragile topology \cite{wang2019, alexandradinata2020, peri2020}. However, the lack of nontrivial boundary modes poses challenges to the identification of fragile TIs in experiments. Recently, it was proposed that fragile topology can be characterized by spectral flow under twisted boundary conditions, which is almost impossible in real materials, but feasible in acoustic systems. The proposed spectral flow was observed in an acoustic crystal through  density of states measurements ({\bf Fig.~\ref{fig6}c}) \cite{peri2020}. This experiment shows that acoustic crystals are suitable platforms to study the exotic physics of fragile topology.

\bigskip
\noindent{\textbf{Non-Hermitian band topology}}
\smallskip

A non-Hermitian system is described by a non-Hermitian Hamiltonian satisfying $\hat{H}\neq\hat{H}^{\dagger}$, with $\dagger$ representing the Hermitian conjugate. Due to the lack of Hermiticity, the eigenvalues of a non-Hermitian system are generally complex, and the corresponding eigenvectors are not guaranteed to be orthogonal. In recent years, people have found that non-Hermiticity has a profound influence on band topology. Here we discuss studies on acoustic non-Hermitian topological systems; for a more comprehensive overview on non-Hermitian topological physics, we refer readers to Refs.~\cite{ghatak2019, bergholtz2021}.

The sources of non-Hermiticity in an acoustic non-Hermitian system are schematically illustrated in {\bf Fig.~\ref{fig1}j}. Losses are ubiquitous in acoustics, and gain and asymmetric couplings can be engineered artificially. In general, studies on acoustic non-Hermitian topology can be divided into two classes.

The first class includes studies in which non-Hermiticity is introduced into originally topological systems \cite{zhu2018, wang2018, zhang2019d, lopez2019, hu2021}. In particular, a 1D acoustic crystal was realized with topological edge states characterized by a nonzero Chern number defined in a 2D synthetic space (synthetic dimensions are discussed in more detail in the next sub-section). Interestingly, when a judicious amount of loss is introduced, topological edge states and an exceptional point can simultaneously exist, leading to unconventional transport phenomena \cite{zhu2018}. Besides, people have also studied the effects of gain and loss on acoustic valley Hall insulators and HOTIs, where interesting effects such as selective amplification and attenuation of topological boundary modes were found \cite{wang2018, zhang2019d, lopez2019, hu2021}.

The second class of studies focuses on non-Hermiticity-induced nontrivial topologies, which include the so-called line-gap topology and point-gap topology \cite{gao2020, gao2021, zhang2021c, zhang2021b, zhang2021e}. A line (or point) gap refers to a line (or point) in the complex frequency plane that does not intersect with any bands \cite{kawabata2019}. Nontrivial line-gapped phases are usually characterized by topological invariants defined by wavefunctions under a biorthogonal basis \cite{brody2013}. In acoustics, nontrivial line gaps can be induced by losses \cite{gao2020, gao2021, zhang2021e}. For example, by deliberately adding an on-site imaginary potential to an acoustic crystal that is gapless, a HOTI with corner states can be obtained \cite{gao2021} ({\bf Fig.~\ref{fig6}d}) \cite{gao2021}. In this case, the on-site losses function like coupling dimerizations. 

Point gaps are associated with novel topological phases that do not have Hermitian counterparts. A nontrivial point gap is characterized by a nonzero winding number, which is defined through complex frequency dispersions rather than wavefunctions \cite{gong2018}. In a finite lattice, such nontrivial point-gap topology manifests as the so-called non-Hermitian skin effect, a phenomenon in which an extensive number of states are localized at the system’s boundaries \cite{yao2018}. The skin modes lead to the breakdown of the Bloch theorem and thus the failure of bulk-boundary correspondence for line-gap topologies that are based on topological invariants computed from Bloch wavefunctions. This problem can be fixed by novel techniques like the generalized BZ method \cite{yao2018}. In acoustics, a straightforward way to realize a nontrivial point gap is to use asymmetric coupling, which can be engineered by introducing losses into the couplers \cite{zhang2021c} or using directional amplifiers \cite{zhang2021b} ({\bf Fig.~\ref{fig6}e}, left). The skin modes can be directly visualized by mapping the field pattern under a point-source excitation ({\bf Fig.~\ref{fig6}e}, right).

Another direction under active exploration in acoustics is the topology associated with exceptional degeneracies (where both eigenvalues and eigenvectors coalesce), such as the discriminant number associated with the eigenvalues and quantized Berry phase accumulated by eigenvectors after encircling the exceptional degeneracies \cite{ding2016, tang2020, tang2021}.

\bigskip
\noindent{\textbf{Topology with synthetic dimensions}}
\smallskip

Another emerging direction is topology with synthetic dimensions, in which some non-spatial degrees of freedom are reinterpreted as additional dimensions, giving access to topological physics in dimensions higher than those of the original system ({\bf Fig.~\ref{fig1}k}). In acoustics, structural parameters are usually used to construct synthetic dimensions. Interested readers may refer to Ref.~\cite{ozawa2019b} for other methods for realizing synthetic dimensions.

Acoustic Weyl physics has been studied in 1D and 2D systems with the help of synthetic dimensions \cite{fan2019, zangeneh2020b, fan2022}. In particular, a 1D layered acoustic crystal was used to realize Weyl points in a 3D space synthesized by a momentum dimension and two structural parameters associated with the layers’ thickness ({\bf Fig.~\ref{fig6}f}) \cite{fan2019}. In this system, the Nielsen-Ninomiya theorem can be bypassed due to the lack of periodicity in the parameter dimensions. A $Z_2$ acoustic Weyl semimetal that supports acoustic pseudospins with Kramers degeneracy has also been realized in a 1D system with two extra synthetic dimensions \cite{zangeneh2020b}.

QH physics has also been studied in 1D acoustic systems with extra synthetic dimensions \cite{ni2019b, long2019, apigo2019, xu2020, chen2021c}. These 1D systems, also called 1D topological pumps, carry a structural modulation term (pumping or phason parameter) that serves as a synthetic momentum. In the 2D synthetic space built from one physical momentum and one pumping parameter, nonzero Chern numbers can be defined. Due to the simplicity of 1D structures and tunability of system parameters, various QH physics, including the spectrum under different parameters and dynamic properties, can be easily accessed in these 1D acoustic systems \cite{ni2019b, long2019, apigo2019, xu2020, chen2021c}. Furthermore, one can extend the 1D topological pumps to 2D and study the physics of the 4D QH effect \cite{chen2021, cheng2021, chen2021d}. In a 2D topological pump, there are two pumping parameters that modulate the 2D system along the $x$ and $y$ directions, respectively ({\bf Fig.~\ref{fig6}g}, left). These two pumping parameters, together with the two physical momenta, form a 4D space where a second Chern number can be defined. When considering a finite sample, acoustic measurements can easily access the boundary modes induced by the nontrivial topology of the bands in the 4D  synthetic space ({\bf Fig.~\ref{fig6}g}, right).

\bigskip
\begin{large}
\noindent\textbf{Outlook}
\end{large}
\smallskip

In just half a decade, the field of topological acoustics has seen  tremendous developments. This is partly because early developments largely overlapped with advances in topological quantum materials ({\bf Fig.~\ref{fig1}l}). We envision that this trend will hold for the next few years, but soon we will need a new roadmap to solve challenges specific to topological acoustics. 

We have to keep in mind that the major motivation for introducing the concept of topology into acoustics is to acquire robust control of sound. The immediate challenge is to understand how to construct various topological phases in acoustics. That is the reason the field of topological acoustics started with emulating electronic topological phases, and now has become an exciting playground for studying various novel topological phases with acoustic waves. In the following, we discuss several opportunities in the field. 

First of all, the emerging directions discussed in this Review, namely, Floquet, fragile, non-Hermitian and synthetic-space topologies, will continue to develop in the next few years. Especially, non-Hermitian topology can produce many unprecedented phenomena in acoustics \cite{zhang2021b, zhang2021c} and may lead to useful devices such as non-Hermitian topoloigcal whispering-gallery cavities \cite{hu2021}. 

One other interesting direction is the exploration of the interplay between topological lattice defects in real space \cite{mermin1979, kosterlitz2017} and band topology in reciprocal space. Historically, the concept of topology was first introduced to describe defects, not band topology \cite{mermin1979, kosterlitz2017}, and the topology for defects and that for bands naturally co-exist in a crystal and shall interact with each other. This has been a topic since the early years of TIs \cite{ran2009, teo2010}, but is rarely studied experimentally due to the lack of suitable platforms. In acoustic structures that can be freely designed and accurately fabricated, topological lattice defects can be easily engineered, and thus the associated physics becomes accessible in the laboratory \cite{wang2021, ye2021, xue2021, lin2021}. 

Another possible direction is nonlinear topological acoustics. Whereas nonlinear phenomena in topological photonics have been studied for a few years \cite{smirnova2020}, in acoustics so far there are only few studies based on spring-mass models \cite{chen2014, pal2018, snee2019, chaunsali2019, darabi2019}. In fact, acoustic waves are intrinsically nonlinear \cite{kinsler2000}. For example,  shock waves, like sonic booms, are nonlinear. Ultrasonic wave propagation for medical use commonly displays nonlinear behaviours. Acoustic levitation is a nonlinear phenomenon as well. Therefore, nonlinear phenomena in topological acoustics should be accessible and fruitful. It might be possible to observe nonlinearity-induced topological phase transitions and topological solitons, similar to the ones found in photonic systems \cite{lumer2013, leykam2016, hadad2016}.

In terms of the connection to fundamental theories in condensed matter physics, a possible direction is to study topological phases under projective symmetries. Most existing topological phases are characterized by either internal or spatial symmetries \cite{chiu2016}, while the effect of a gauge symmetry is less explored. Recently, it has been shown that projectively represented lattice symmetries under a gauge field can lead to novel topological phenomena \cite{zhao2020, zhao2021, shao2021}. Acoustic systems, in which different gauge configurations can be realized with proper negative couplings \cite{xue2021b, li2021}, offer an experimental avenue to explore projectively enriched topological physics. When it comes to acoustic topological semimetals, many types of band degeneracies remain unexplored, including higher-order Weyl and Dirac points, high-fold nodal points and different types of nodal lines. Moreover, acoustic systems are practical platforms to engineer gauge fields for band degeneracies from structural tuning. So far, however, artificial gauge fields have only been constructed for Dirac points in 2D \cite{wen2019} and Weyl points in 3D \cite{peri2019}. Beyond acoustic topological semimetals, artificial gauge fields themselves are powerful tools to manipulate classical waves, even when the underlying effective magnetic field is zero. Studies of spatially uniform artificial gauge fields have already revealed some novel wave phenomena such as gauge-field-induced negative refraction and waveguiding \cite{yang2021b}. Another potential direction is non-Abelian band topology in acoustic crystals, which may offer insights to new topological physics in multi-band systems \cite{jiang2021, wang2022}. As the disorder in acoustic systems can be precisely engineered, it would also be interesting to investigate the interplay between disorder and topology in acoustics, which could enable counterintuitive disorder-induced topological Anderson insulators \cite{zangeneh2020c}. 

A lot of efforts need to be devoted to pushing topological acoustics into real-world applications. We now know how to construct topological phases for sound, but there are still challenges for practical applications. First of all, most studies in topological acoustics are limited to airborne sound, whereas elastic waves \cite{huber2016, ma2019}, which contain both transverse and longitudianl modes, are much less studied. How to integrate the transverse and longitudinal modes is still an open question, especially in 3D geometries. Secondly, most studies have been performed in audible frequency ranges with long wavelengths. It is still unclear how to push the working frequency into the ultrasound and even hypersound ranges. These are the frequency ranges of relevance to medicine and modern industry.

The good news is that any improvement brought by the robustness of topological acoustic waves will fit into the broad frequency spectrum of acoustics, which has plenty of potential applications. In future acoustic applications, topological corner, edge or surface states could be used to realize devices robust against defects and disorder, such as acoustic lasers, absorbers, waveguides, multiplexers, filters, resonators, sensors and antennas. Topological acoustic states could also significantly impact acoustic information technologies, in which they will lay the foundations for robust information transfer, encoding and decoding, processing and storage. Applying topological acoustics to surface acoustic waves may inspire the next generation of surface acoustic wave technologies, which could process radio-frequency signals in portable devices such as mobile phones. Finally, as topology is a fundamental degree of freedom in acoustics, it has the potential to reshape all acoustic-wave-based technologies, such as underwater communications, ultrasound imaging, noise control and many others.

\bigskip

\noindent\textbf{Acknowledgements}

\noindent H.X. and B.Z. acknowledge support from National Research Foundation Singapore Competitive Research Program No. NRF-CRP23-2019-0007, and Singapore Ministry of Education Academic Research Fund Tier 3 under Grant No. MOE2016-T3-1-006 and Tier 2 under Grant No. MOE2019-T2-2-085. Y.Y acknowledges support from the National Natural Science Foundation of China (NNSFC) under Grant No. 62175215.\\

\noindent\textbf{Author contributions}

\noindent All authors contributed significantly to all aspects of the article.\\

\noindent\textbf{Competing interests}

\noindent The authors declare no competing interests.

\begin{figure*}
  \centering
  \includegraphics[width=0.85\textwidth]{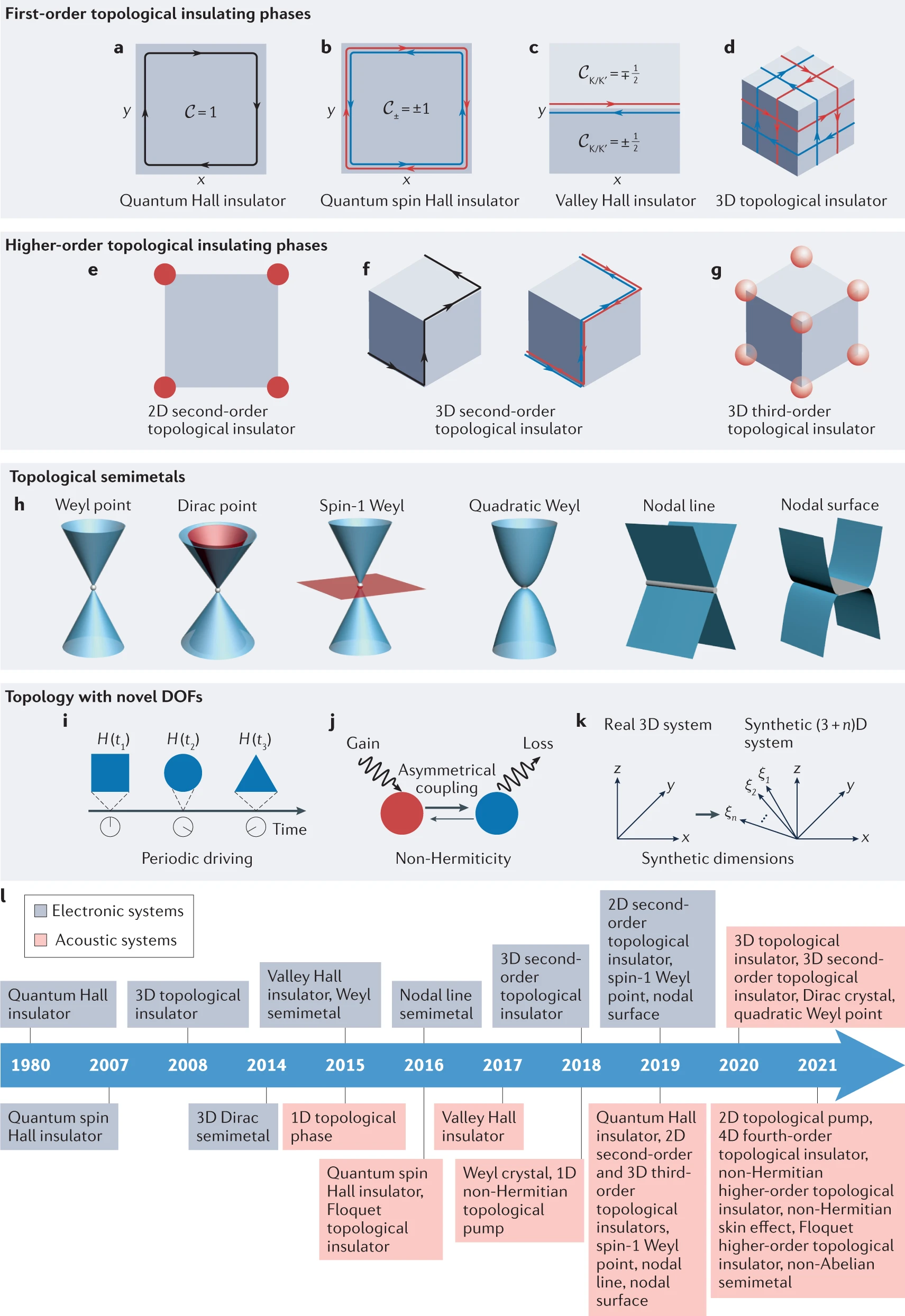}
\end{figure*}
\begin{figure*}
  \caption{$|$ \textbf{Summary of different types of topological phases.} \textbf{a} $|$ Schematic of a quantum Hall insulator. Chiral edge modes propagate unidirectionally in a finite sample. $\mathcal{C}$ denotes the Chern number. \textbf{b} $|$  Schematic of a quantum spin Hall insulator, where spin-up (red) and spin-down (blue) states propagate in opposite directions along the edges. $\mathcal{C_{+}}$ and $\mathcal{C_{-}}$ denote the Chern numbers for spin-up and spin-down states, respectively. \textbf{c} $|$ Schematic of a valley Hall insulator. Valley kink states (red and blue) locked to the K and $\text{K}'$ valleys, respectively, emerge at the interface between two domains with opposite valley-Chern numbers (denoted by $\mathcal{C_{\text{K}}}$ and $\mathcal{C_{\text{K}'}}$).  \textbf{d} $|$ Schematic of a 3D topological insulator with spin-polarized topological surface states. \textbf{e} $|$ Schematic of a 2D second-order topological insulator. Topological corner modes (red) emerge at corners. \textbf{f} $|$ Schematic of 3D second-order topological insulators. Left: chiral hinge modes propagate along edges. Right : spin-up and spin-down states propagate in opposite directions along edges.  \textbf{g} $|$ Schematic of a 3D third-order topological insulator. Topological corner modes (red) emerge at corners. \textbf{h} $|$ Plots of various band degeneracies, including a Weyl point,  a Dirac point, a three-fold Weyl point, a quadratic Weyl point, a nodal line and a nodal surface. \textbf{i} $|$ Schematic of a Floquet (periodically driven) system, whose Hamiltonian $H$ varies periodically (illustrated by different shapes) with time. \textbf{j} $|$ Schematic of a non-Hermitian system of two coupled resonators. The non-Hermiticity originates from gain (red), loss (blue), or asymmetric coupling. \textbf{k} $|$ Schematic of a system with synthetic dimensions. A real 3D system (left) can be generalized to a synthetic $(3+n)$D system (right) by adding several system parameters (denoted by $\xi_1,\cdots,\xi_n$) as synthetic dimensions. \textbf{l} $|$ Timeline of experimental realization of various topological phases in electronic and acoustic systems. The corresponding references are as follows. Electronic systems: quantum Hall insulator \cite{klitzing1980}, quantum spin Hall insulator \cite{konig2007}, 3D topological insulator \cite{hsieh2008}, 3D Dirac semimetal \cite{liu2014}, valley Hall insulator \cite{ju2015}, Weyl semimetal \cite{xu2015, lv2015}, nodal line semimetal \cite{bian2016}, 3D second-order topological insulator \cite{schindler2018}, 2D second-order topological insulator \cite{kempkes2019}, spin-1 Weyl point \cite{rao2019, sanchez2019, takane2019} and nodal surface \cite{fu2019}. Acoustic systems: 1D topological phase \cite{xiao2015a}, quantum spin Hall insulator \cite{he2016}, Floquet topological insulator \cite{peng2016}, valley Hall insulator \cite{lu2017}, Weyl crystal \cite{li2018b, ge2018}, 1D non-Hermitian topological pump \cite{zhu2018}, quantum Hall insulator \cite{ding2019}, 2D second-order topological insulator \cite{xue2019a, ni2019, zhang2019a}, 3D third-order topological insulator \cite{xue2019b, zhang2019b, weiner2020}, spin-1 Weyl point \cite{yang2019}, nodal line \cite{deng2019, qiu2019}, nodal surface \cite{yang2019b, xiao2020}, 3D topological insulator and 3D second-order topological insulator \cite{he2020}, Dirac crystal \cite{cheng2020, cai2020}, quadratic Weyl point \cite{he2020b}, 2D topological pump \cite{chen2021, cheng2021, chen2021d}, 4D fourth-order topological insulator \cite{chen2021b}, non-Hermitian higher-order topological insulator \cite{gao2021}, non-Hermitian skin effect \cite{zhang2021c, zhang2021b}, Floquet higher-order topological insulator \cite{zhu2020} and non-Abelian semimetal \cite{jiang2021}. DOFs: degrees of freedom}
  \label{fig1}
\end{figure*}

\begin{figure*}
\centering
\includegraphics[width=0.9\textwidth]{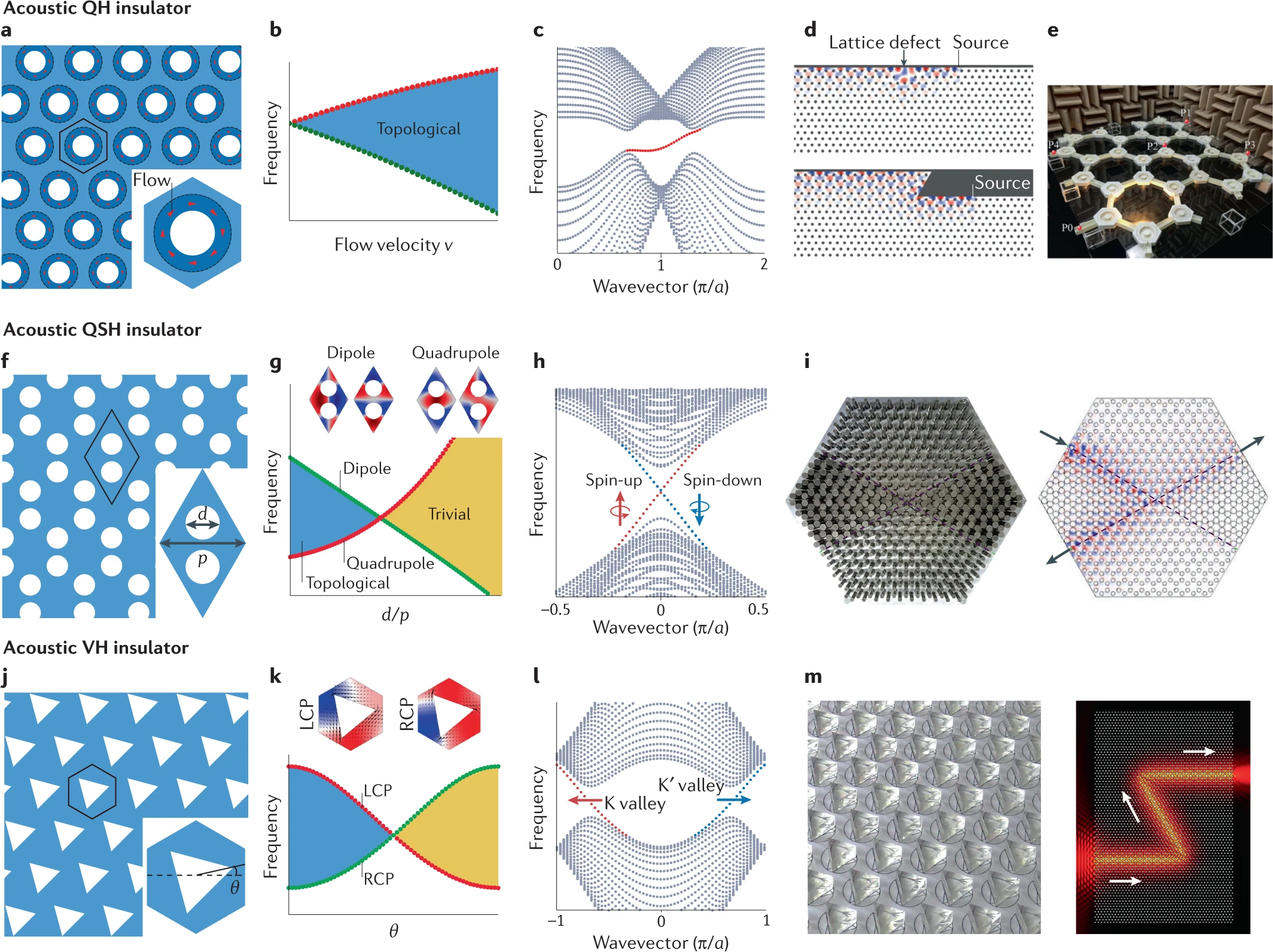}
\caption{$|$ \textbf{Acoustic analogues of 2D topological insulating phases.} \textbf{a} $|$ Schematic of an acoustic QH insulator lattice. The white regions are rotating rigid solids, and the dark blue regions are filled with fluids that circulate as indicated by the red arrows. The rest of the lattice is air. \textbf{b} $|$ Frequency splitting at the Dirac point (a two-fold linear degenerary) as the flow velocity increases from zero. \textbf{c} $|$ Edge dispersion. The chiral edge state is denoted by the red line. \textbf{d} $|$ Simulated sound field distributions of the chiral edge state in a lattice with a cavity defect (top) and in a lattice with a path bend (bottom). \textbf{e} $|$ Sample realizing an acoustic QH insulator. \textbf{f} $|$ Schematic of an acoustic quantum spin Hall (QSH) insulator that consists of an array of rigid rods (white) in air (light blue) in a honeycomb lattice. $p$ and $d$ denote the lattice constant and rod diameter, respectively. \textbf{g} $|$ Topological phase transition. In the plot, the blue and yellow regions represent topologically nontrivial and trivial phases, respectively. The green and red lines are the dipole and quadrupole modes, respectively. The topological phase transition happens at $d/p=0.4536$. The insets at the top show the dipole and quadrupole modes at the centre of the Brillouin zone (BZ). \textbf{h} $|$ Edge dispersion. Red and blue dotted lines indicate the pseudospin-up and pseudospin-down topological edge states. \textbf{i} $|$ An acoustic QSH insulator (left) and the simulated field distribution (right) when sound is launched from the top left port. The sound wave cannot come out from the lower right port due to pseudospin conservation. \textbf{j} $|$ Schematic of an acoustic valley Hall (VH) insulator consisting of an array of rigid triangular scatterers (white) in air (light blue) in a triangular lattice. $\theta$ denotes the orientation angle of the scatterers. \textbf{k} $|$ Topological phase transition. The blue and orange regions are topologically distinct. The green and red lines represent left-handed and right-handed circular polarization modes, respectively. The insets at the top show left-handed (LCP) and right-handed circular polarization (RCP) modes at the BZ corners. \textbf{l} $|$ Edge dispersion. The red and blue dotted lines indicate the kink states locked at the K and $\text{K}’ $ valleys, respectively. \textbf{m} $|$ An acoustic VH insulator (left) and the simulated field distribution when sound is launched from the left port. The arrows show the sound propagation direction. Panel \textbf{e} adapted with permission from Ref.~\cite{ding2019} (Ding et al.). Panel \textbf{d} adapted with permission from Ref.~\cite{yang2015} (Yang et al.). Panel \textbf{i} adapted with permission from Ref.~\cite{he2016} (He et al.). Panel \textbf{m} adapted with permission from Ref.~\cite{lu2017} (Lu et al.).}
\label{fig2}
\end{figure*}

\begin{figure*}
  \centering
  \includegraphics[width=0.9\textwidth]{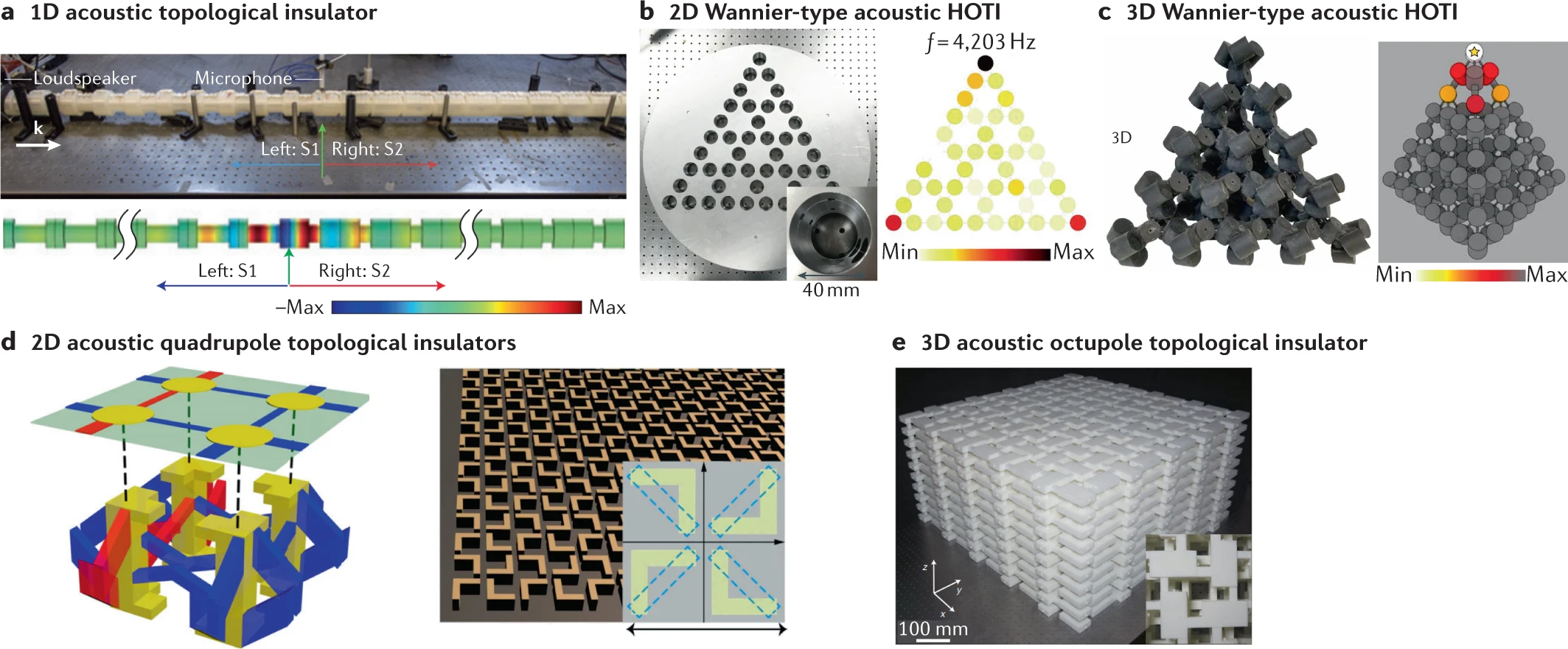}
  \caption{$|$ \textbf{Acoustic topological phases with quantized dipole and multipole moments.} \textbf{a} $|$ A 1D acoustic structure consisting of a nontrivial lattice, S1, and a trivial lattice, S2 (top), and simulated field pattern of the topological mode localized at the interface between the two lattices (bottom). The color represents acoustic pressure.  \textbf{b} $|$ A kagome lattice sample (left), and the measured local density of states at 4203 Hz, which exhibits localization at the three corners (right).  \textbf{c} $|$ A pyrochlore lattice sample (left), and the measured acoustic field distribution at corner modes' frequency (right). The star denotes the source which excites one of the corner modes. \textbf{d} $|$ Unit cell of an acoustic quadrupole topological insulator based on coupled acoustic resonators (left). Red and blue indicate channels that enable positive and negative couplings, respectively. A 2D nonsymmophic sonic crystal that realizes a quadrupole topological insulator (right). The inset shows the unit cell.  \textbf{e} $|$ An acoustic octupole topological insulator. Panel \textbf{a} adapted with permission from Ref.~\cite{xiao2015a} (Xiao et al.). Panel \textbf{b} adapted with permission from Ref.~\cite{xue2019a} (Xue et al.). Panel \textbf{c} adapted with permission from Ref.~\cite{weiner2020} (Weiner et al.). Panel \textbf{d} left adapted with permission from Ref.~\cite{qi2020} (Qi et al.), panel d right from Ref.~\cite{zhang2020} (Zhang et al.). Panel \textbf{e} adapted with permission from Refs.~\cite{xue2020} (Xue et al.).}
  \label{fig3}
\end{figure*}

\begin{figure*}
  \centering
  \includegraphics[width=0.9\textwidth]{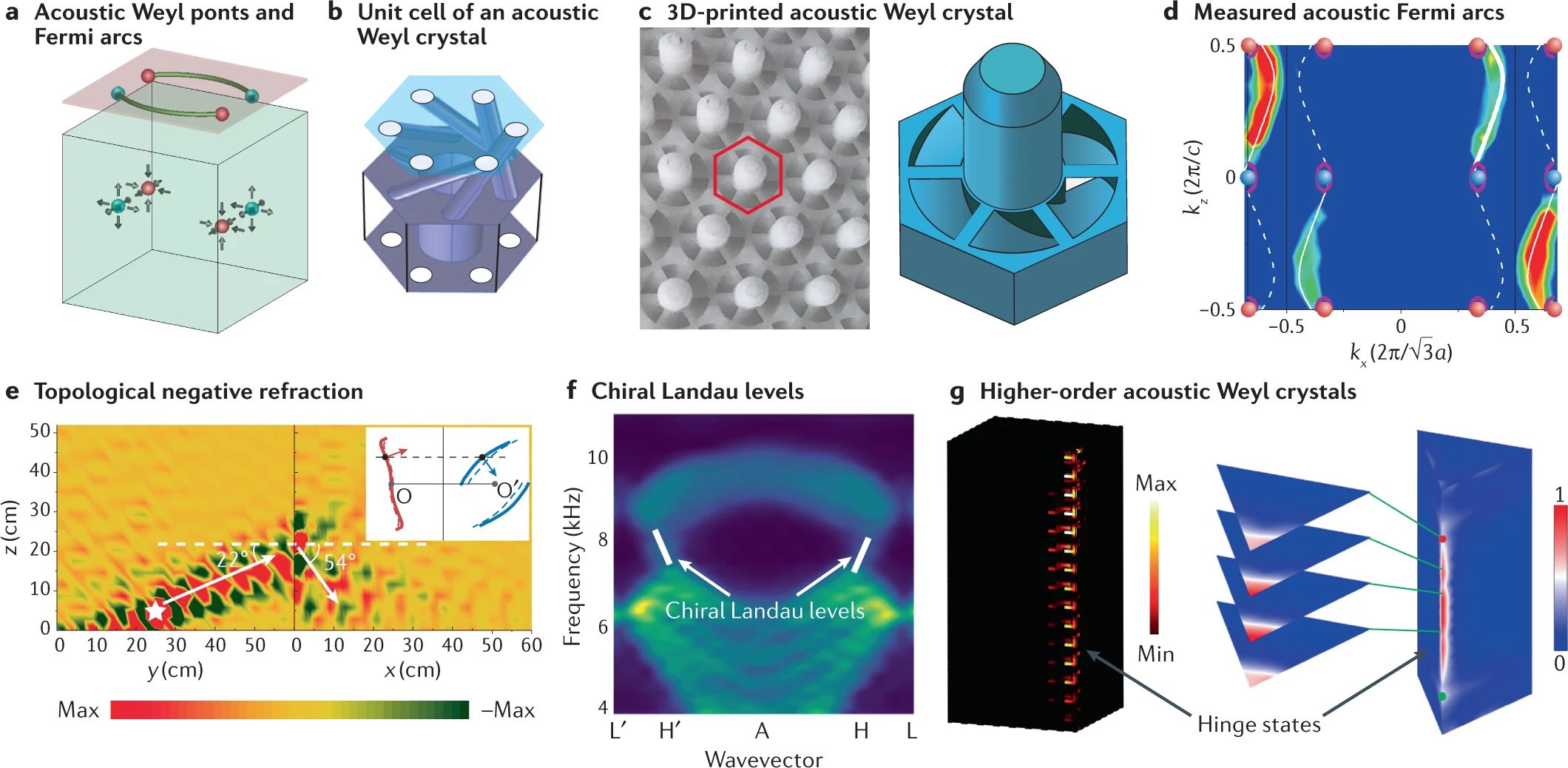}
  \caption{$|$ \textbf{Acoustic Weyl crystals.} \textbf{a} $|$ Fermi arcs from acoustic Weyl points. The surface projections of two oppositely charged Weyl points are connected via non-closed Fermi arcs. \textbf{b} $|$ Unit cell of an acoustic Weyl crystal. \textbf{c} $|$ Picture (left) and unit cell (right) of a 3D-printed acoustic Weyl crystal. \textbf{d} $|$ Measured acoustic Fermi arcs on the front surface. The solid and dashed lines show simulated Fermi arcs on the front and back surfaces. The red and blue dots denote the projections of Weyl points with opposite charges. \textbf{e} $|$ Topological negative refraction. By judiciously engineering the Fermi arcs on two adjacent surfaces, topological negative refraction without reflection can be obtained. The left and right regions represent the measured field distributions at the $yz$ and $xz$ planes, respectively. The star denotes the position of the source. The arrows indicate the propagation direction of the surface acoustic wave. The inset shows the Fermi arcs on two surfaces, where the letters `$O$'and `$O'$' stand for the origin of momentum space, and the arrows represent the direction of group velocity. \textbf{f} $|$ Measured chiral Landau levels in an inhomogeneous acoustic Weyl crystal.  \textbf{g} $|$ Simulated acoustic field distribution when a source is placed at the hinge of a higher-order acoustic Weyl crystal (right). Measured hinge state field distributions for various $k_z$, with the red and blue dots the projections of Weyl points with opposite charges (right). Panel \textbf{b} adapted with permission from Ref.~\cite{xiao2015b} (Xiao et al.). Panels \textbf{c} and \textbf{d} adapted with permission from Ref.~\cite{li2018b} (Li et al.). Panel \textbf{e} adapted with permission from Ref.~\cite{he2018} (He et al.). Panel \textbf{f} adapted with permission from Ref.~\cite{peri2019} (Peri et al.). Panel \textbf{g} left adapted with permission from Refs.~\cite{luo2021} (Luo et al.), right from Refs.~\cite{wei2021} (Wei et al.).}.
  \label{fig4}
\end{figure*}

\begin{figure*}
  \centering
  \includegraphics[width=0.9\textwidth]{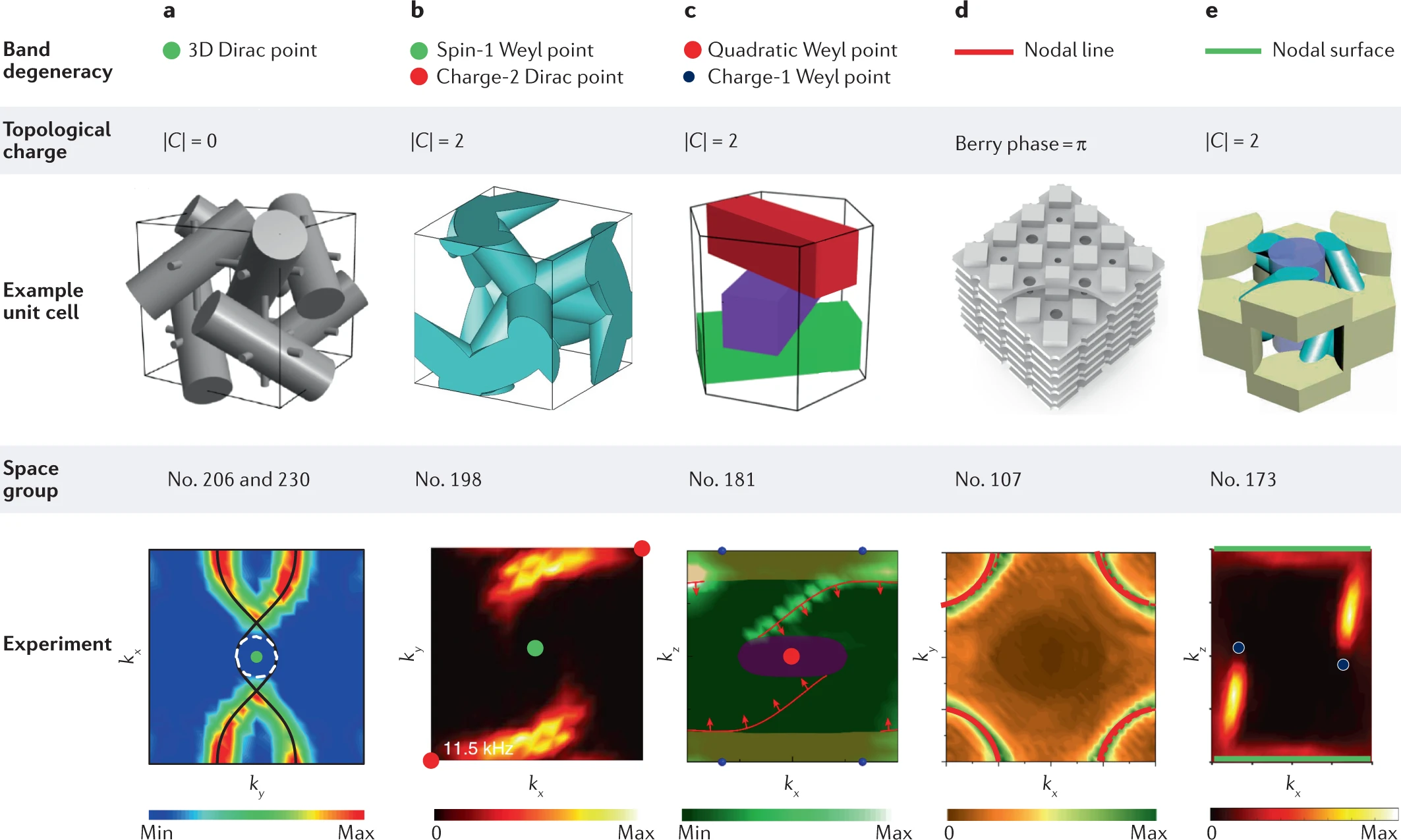}
  \caption{$|$ \textbf{Acoustic topological semimetals beyond conventional Weyl phases.} \textbf{a} $|$ 3D Dirac points with quad-helicoid surface states. The green dot represents the projection of the 3D Dirac points. \textbf{b} $|$ Spin-1 Weyl point and charge-2 Dirac point with double helicoid Fermi-arc surface states. The green and red dots represent the projection of the spin-1 Weyl and the charge-2 Dirac points, respectively. \textbf{c} $|$ Quadratic Weyl point with double Fermi arcs. The red lines are the numerically calculated Fermi arcs. The arrows represent the direction of the group velocity. The blue and red dots indicate the charge-1 and quadratic Weyl points, respectively. \textbf{d} $|$ Topological nodal ring. The red curves represent the nodal ring pinned at the $k_z=0$ plane. \textbf{e} $|$ Topologically charged nodal surfaces with double Fermi arcs. The Fermi arcs connect the charge-2 nodal surfaces to two charge-1 Weyl points. The green lines represent the projection of the nodal surface. The fifth row in each panel shows the dispersion obtained from Fourier transform of measured acoustic field. The third row in panel \textbf{a} adapted with permission from Ref.~\cite{cheng2020} (Cheng et al.). The fifth row in panel \textbf{a} adapted with permission from Ref.~\cite{cai2020} (Cai et al.). The third and fifth rows in panel \textbf{b} adapted with permission from Ref.~\cite{yang2019} (Yang et al.). The third and fifth rows in panel \textbf{c} adapted with permission from Ref.~\cite{he2020b} (He et al). The third and fifth rows in panel \textbf{d} adapted with permission from Ref.~\cite{deng2019} (Deng et al.). The third row in panel \textbf{e} adapted with permission from Ref.~\cite{xiao2020} (Xiao et al.). The fifth row in panel \textbf{e} adapted with permission from Ref.~\cite{yang2019b} (Yang et al.).}
  \label{fig5}
\end{figure*}

\begin{figure*}
  \centering
  \includegraphics[width=\textwidth]{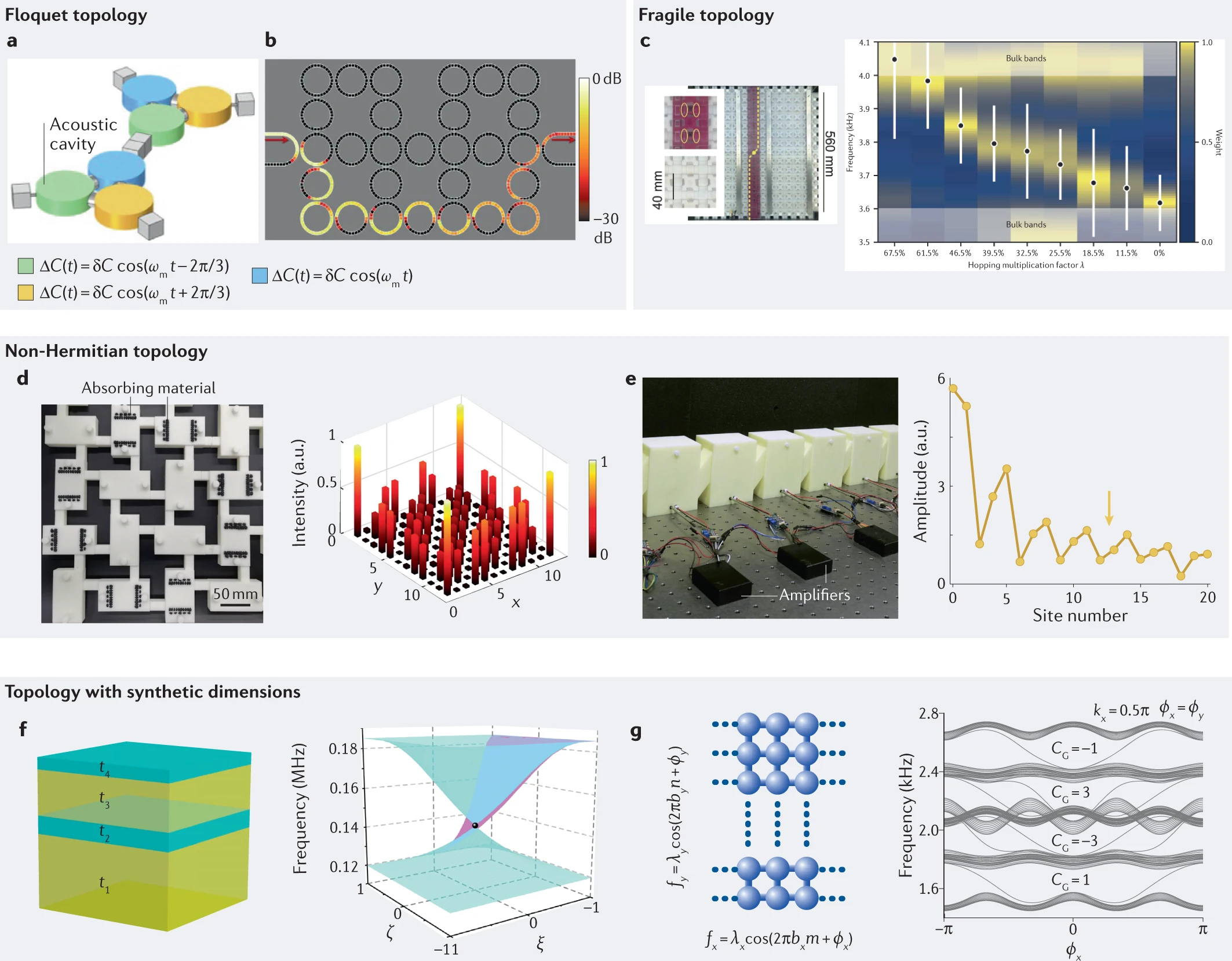}
  \caption{$|$ \textbf{Novel acoustic topological phases.} \textbf{a} $|$ Acoustic Floquet topological insulator (TI). Each unit cell consists of two trimers made of three coupled acoustic cavities each. The acoustic capacitance ($C$) of each cavity is periodically modulated in a rotating fashion, as detailed by the equation for the capacitance of each cavity. Here $\delta C$ and $\omega_m$ are the modulation amplitude and frequency, respectively. \textbf{b} $|$ Acoustic anomalous Floquet TI consisting of judiciously designed coupled ring resonators that support pseudospin-polarized topological edge states. \textbf{c} $|$ A fragile acoustic TI sample with a twisted boundary condition (left). Here two lattices are connected by a thin channel (the purple area). By inserting obstructions of different sizes along the yellow dashed line, different multiplication factor of the hoppings in the purple area can be implemented. The insets show the details of the bulk lattice (bottom) and the boundary area (top). Measured local density of states at sample symmetry center for different values of hopping multiplication factor (right), which reveals the spectral flow under the twisted boundary condition. \textbf{d} $|$ A non-Hermiticity-induced higher-order acoustic TI (left). Some acoustic cavities are filled with black absorbing material to enhance losses. The intensity profiles measured across the sample are shown in the graph on the right. \textbf{e} $|$ Acoustic crystal with non-Hermitian topology (left). The acoustic amplifier connecting two adjacent cavities provides asymmetric coupling. Measured field intensities (right). All modes are localized at the left boundary. The arrow denotes the position of the source. \textbf{f} $|$ Schematic of a multi-layer structure, where $t_{1}$--$t_{4}$ denote the layers' thickness (left). Dispersion for this structure in a synthetic space (right). Blue and pink bands correspond to bulk and surface states, respectively. The black dot denotes the Weyl point. The parameters $\xi$ and $\zeta$, serving as synthetic dimensions, are defined as $\xi=(t_1-t_3)/(t_1+t_3)$ and $\zeta=(t_2-t_4)/(t_2+t_4)$. \textbf{g} $|$ Schematic of a 2D lattice with modulations along the $x$ and $y$ directions (left). The on-site frequencies are modulated according to the formulas given in the plot, with $\lambda_{x,y}$ the modulation amplitude, $b_{x,y}$ the modulation frequency, and $m$ and $n$ the site indices along the $x$ and $y$ directions. The phase parameters $\phi_x$ and $\phi_y$ are two synthetic dimensions. Plot of the calculated eigenfrequencies of the system with finite sizes in the $x$ and $y$ directions against the pumping parameter $\phi_x$ (right). $C_G$ denotes the second Chern number for each gap, which induces gapless boundary modes. Panel \textbf{a} adapted with permission from Ref.~\cite{fleury2016} (Fleury et al.). Panel \textbf{b} adapted with permission from Ref.~\cite{peng2016} (Peng et al.). Panel \textbf{c} adapted with permission from Ref.~\cite{peri2020} (Peri et al.). Panel \textbf{d} adapted with permission from Ref.~\cite{gao2021} (Gao et al.). Panel \textbf{e} adapted with permission from Ref.~\cite{zhang2021b} (Zhang et al.). Panel \textbf{f} adapted with permission from Ref.~\cite{fan2019} (Fan et al.). Panel \textbf{g} adapted with permission from Ref.~\cite{chen2021} (Chen et al.).}
  \label{fig6}
\end{figure*}

\end{document}